\documentstyle[11pt,paspconf,epsfig]{article}
\def\ltsima{$\; \buildrel < \over \sim \;$}
\def\lsim{\lower.5ex\hbox{\ltsima}}
\def\gtsima{$\; \buildrel > \over \sim \;$}
\def\gsim{\lower.5ex\hbox{\gtsima}}

\begin{document}

\title{Quasi--Thermal Comptonization and GRBs}

\author{Gabriele Ghisellini}
\affil{Osservatorio Astron. di Brera, Via Bianchi 46, I-23807 Merate, Italy}

\author{Annalisa Celotti}
\affil{S.I.S.S.A., Via Beirut 2/4, I--34014 Trieste, Italy}

\begin{abstract}
Quasi--thermal Comptonization is an attractive alternative to
the synchrotron process to explain the spectra of GRBs, even if we 
maintain other important properties of the internal shock scenario, 
implying a compact emitting region and an equipartition magnetic field. 
Photon--photon absorption and electron--positron pairs can play a crucial 
role: this process may lock the effective temperature in a narrow range
and may be the reason why burst spectra have high energy cut--offs
close to the rest mass--energy of the electron.
If the progenitors of GRB are hypernovae, the circum--burst matter
is dominated by the wind of the pre--hypernova star. 
The presence of this dense material has strong effects on the
generation of the radiation of the burst and its afterglow.
\end{abstract}

% Keywords should be included, but they are not printed in the hardcopy.

\keywords{emission mechanisms: Comptonization --- Hypernovae ---
electron--positron pairs}

\section{Introduction}

In Celotti \& Ghisellini (this volume) we argue that 
interpreting the burst emission as synchrotron radiation faces
some severe problems.
This justifies the search for alternatives. 
Here we will argue that a valid alternative is quasi--thermal
Comptonization.
This has already been proposed to explain the burst emission by
Liang (1997) and Liang et al. (1997), who required a relatively weakly 
magnetized (magnetic field $B\sim 0.1$ Gauss) and large 
($R\sim 10^{15}$ cm) emitting region.
These values of the physical parameters contrast with the ones
advocated by the ``standard internal shock" scenario as it has been
developing to explain the structured GRB light curve and its fast 
variability, which requires a compact ($R\sim 10^{13}$ cm) and 
magnetized ($B\sim 10^5$ Gauss) emitting region
(Rees \& M\'esz\'aros 1992; Rees \& M\'esz\'aros 1994; Sari \& Piran 1997).

More recently, we (Ghisellini \& Celotti 1999) have proposed again
the quasi--thermal Comptonization scenario, but using the very same
physical parameters as in the internal shock picture.
The only (important) difference concerned
the timescale of particle acceleration.
Instead of considering it instantaneous, we considered that the 
particles can be re--accelerated for the entire duration of the
shell--shell interaction, and therefore the acceleration timescale
can last for $\Delta R^\prime /c$, where $\Delta R^\prime$ 
is the shell width as measured in the comoving frame.

In this case the typical electron energy is dictated by the balance between 
the heating and the cooling rate: assuming that the bulk Lorentz factor
of one shell in the comoving frame of the other is $\Gamma^\prime$, we obtain:
\begin{equation}
{(\Gamma^\prime-1) n^\prime_p m_p c^2 \over \Delta R^\prime/c} \, =\,
{4\over 3} n_e^\prime \sigma_T c (\gamma^2-1) U
\end{equation}
where $n_p^\prime$ and $n_e^\prime$ are the comoving densities of protons
and leptons, respectively, and $U$ is the total (radiative plus magnetic)
energy density.
The resulting particle distribution may well be different from a perfect
Maxwellian, but it has in any case a narrow energy width, and a
meaningful mean energy may be defined.
It may also be possible that a high energy tail (possibly a steep power law)
is present, producing a tail of high frequency radiation (see e.g. Stern 1999).

With such low values of the typical $\gamma$, the produced cyclo--synchrotron 
radiation is self--absorbed and the corresponding power is orders of
magnitudes lower than the observed burst power and at much lower typical 
frequencies. 
This self absorbed radiation is nevertheless important, because it provides
the seed photons to be scattered at high energies.

In this paper we will show why multiple Compton can provide typical
spectral slopes in agreement with observations, and some ideas on how
it is possible to have spectral cut--offs at the observed energies.
Finally we will present some considerations on the hypernova scenario,
based on the fact that the pre--hypernova star necessarily has a 
strong wind, which makes the the circum--burst surroundings very dense.
The effects of this wind will be considered and discussed.

\section{Quasi thermal--Comptonization}

As mentioned above, the particle distribution may not be a perfect Maxwellian,
but it can nevertheless have a well defined mean energy, which
can correspond to an effective temperature.
Let us then introduce a dimensionless effective temperature
$\Theta^\prime \equiv kT^\prime /(m_ec^2)$, measured in the comoving frame.
Assume also that all particles in the shell,
of optical depth 
$\tau^\prime\equiv \sigma_T n_e^\prime \Delta R^\prime$,
partecipate to the burst emission.
The interaction between the shells
is at a distance $R_i=10^{13}R_{i,13}$ cm from the center, and the
shell width is $\Delta R^\prime \sim R/\Gamma$.

\subsection{Seed photons}

As long as the typical $\gamma$ factor of the emitting electron is low, the 
cyclo--synchrotron radiation is self absorbed, and the corresponding
spectrum resembles a blackbody, peaking at the self-absorption frequency
$\nu_T^\prime$, which is a strong function of the temperature.
Interpolating numerical results, Ghisellini \& Celotti (1999) obtain
$\nu_{T}^\prime \sim 2.75\times 10^{14}(\Theta^\prime)^{1.191}$ Hz,
which holds for $0.1 \lsim \Theta^\prime \lsim 3$.
This gives, for $B \sim 10^5$ Gauss and $\tau_T\sim 1$, 
$\nu_{T}^\prime \sim 2.75\times 10^{14} (\Theta^\prime)^{1.191}$ Hz.

The corresponding comoving self--absorbed luminosity is
\begin{equation}
L^\prime_s \, \sim {8\pi \over 3} \, m_e R^2\Theta^\prime
(\nu_T^\prime)^3 \, \sim 7.6\times 10^{41} \Theta^\prime R_{13}^2
(\nu_{T,14}^\prime)^3 \,\, {\rm erg~s^{-1}}
\end{equation}
The same electrons will scatter these photons through multiple scatterings,
in order to emit the burst luminosity.
We can define a generalized Comptonization parameter as
\begin{equation}
y\, \equiv\, 4\tau\Theta^\prime (1+\tau)(1+4\Theta^\prime)
\end{equation}
For values of $y$ larger than unity the final spectrum amplifies the 
synchrotron power by the factor $e^y$: values of $y$ around 10--13 are needed
to produce an intrinsic
Compton power $L_c^\prime \sim 10^{46}$ erg s$^{-1}$ starting from
a synchrotron power $L_s^\prime\sim 10^{41}$ erg s$^{-1}$.

\subsection{A preferred slope: $\nu^0$}

Thermal Comptonization has been extensively studied in the past years
to explain the high energy spectra of galactic black hole candidates
and radio--quiet AGNs (see e.g. Pozdnyakov, Sobol, \& Sunyaev 1983).
The fractional energy amplification of the scattered photons, at each
scattering, is $A\sim 1+4\Theta^\prime+16 (\Theta^\prime)^2$.
When $\tau$ is significantly larger than unity, almost all photons
undergo several scatterings, and in the Compton spectrum of each order we 
therefore have the same number of photons, of mean frequency $\nu_i$
and distributed in a range $\Delta \nu_i\sim \nu_i$ of frequencies.
The {\it escaping} photons are a fraction $\sim 1/\tau$ of the
ones contained in each spectrum.
We therefore have that, in a $\nu$--$F_\nu$ plot, the spectrum of the 
escaping photons is flat ($F_\nu\propto \nu^0$) up to $\sim \Theta^\prime$, 
where photons have the same energies of the leptons.
At these energies a Wien peak forms 
($F_\nu\propto \nu^3 \exp(-h\nu/kT^\prime)$, 
whose importance depends on the value of $\tau$ and $\Theta^\prime$.
In this case an increased (decreased) $\tau$ and/or $\Theta^\prime$
make the Wien peak to become
more (less) dominant, and at the same time they decrease (increase)
the normalization of the power law part of the spectrum, 
but {\it they do not change its slope}.

\subsection{Importance of pairs and feedbacks}

The production of electron--positron pairs would surely be efficient for 
intrinsic compactnesses $\ell^\prime>1$
\footnote{We define the compactness as 
$\ell = \sigma_T L \Delta R^\prime /(m_ec^3 R^2)$. 
See Celotti \& Ghisellini, this volume, for more details.}, 
and would on one hand increase the optical depth, and on the other 
acts as a thermostat, by maintaining the temperature in a narrow range.  
Detailed time dependent studies of the optical depth and temperature 
evolution for a rapidly varying source have not yet been pursued.  
Results concerning a steady source in pair equilibrium indicate 
that for $\ell^\prime$ between 10 and $10^3$ the maximum equilibrium 
temperature is of the order of 30--300 keV (Svensson 1982, 1984), 
if the source is pair dominated (i.e. the density of pairs
outnumbers the density of protons). 
Indeed we expect in this situation to be close to pair equilibrium, 
as this would be reached in about a dynamical timescale 
(i.e. in $\Delta R^\prime/c$), but note that the quoted numbers 
refer to a perfect Maxwellian particle distribution.
If an high energy tail is present, more photons are created above 
the threshold for photon--photon pair production with respect to the case
of a pure Maxwellian, and thus pairs become important for 
values of $\Theta^\prime$ lower than in the completely thermal case 
(see Stern 1999; Coppi 1999; Stern, this volume).

An `effective' temperature of $kT^\prime\sim$ 50 keV
($\Theta^\prime\sim 0.1$) and $\tau_T\sim 4$ dominated by
pairs, can be a consistent solution giving $y\sim 11$.  
See also below for an effect which could considerably enhance the 
compactness of the emitting region, and therefore its pair density.

\section{The high energy cut--off}

With $\Theta^\prime\sim 0.1$, and $\Gamma\sim 100$, the observed high energy 
cutoff lies at $E_c \sim 10 \Theta^\prime_{-1}\Gamma_2/(1+z)$ MeV.  
This value is somewhat larger than what is typically observed.
However, there are a number of effects which may be potentially important,
and that can lower this value.
One is that the entire system is highly time--dependent, and the 
time--evolution is in the sense of a cooling of the leptons, 
which will then produce a time--averaged spectral energy
cut--off lower than a few MeV.
On the other hand, if the observed value of $300$ keV is really typical,
and not biased by selection effects introduced by triggering criteria and
detector response energies (see e.g. Lloyd \& Petrosian 1999, and 
Petrosian, this volume), 
we ought to look for a very robust explanation.

\subsection{``Brainerd break"}

Brainerd (1994) linked the typical high energy cut--off of GRBs to
the effect of down--scattering: photons with energies much 
larger than $m_ec^2$ pass undisturbed through a scattering medium 
because of the reduction with energy of the Klein Nishina
cross section, while photons with energies just below $m_ec^2$ 
interact, and their energy after the scattering is reduced.
The net effect is to produce a ``downscattering hole" in the spectrum,
between $\sim m_ec^2/\tau^2$ and $\sim \tau m_ec^2$.
The attractive feature of this model is that the cut--off energy is
associated with the rest mass-energy of the electron.
The difficulty is that a significant part of the power originally 
radiated by the burst goes into heating (by the Compton process) 
of the scattering electrons.

\subsection{Pair production break}

If some scatterings take place between the burst photons and
some external medium at rest, there may be a very efficient process
which modifies the emergent burst radiation, namely pair production.
Assume in fact that the external medium has an optical depth $\tau_{ext}$
in a region close to where the burst radiation originates (i.e. between
$R_i$ and $2R_i$).
This material will scatter back a fraction $\tau_{ext}L$ of the 
burst power, corresponding to a compactness
\begin{equation}
\ell_{ext}\, \approx \, {\sigma_T \tau_{ext} L \over R_i m_ec^3}
\end{equation}
If we require that the primary spectrum is not modified by photon--photon 
absorption, the optical depth of the scattering matter and its density must be
\begin{equation}
\tau_{ext} \, < \, 3.7\times 10^{-9} {R_{13}\over L_{50}} 
\, \to \, n_{ext} \, <\, {5.5\times 10^{2} \over L_{50}}\, \, 
{\rm cm^{-3}}
\end{equation}
As can be seen, the requirement on the density of the external matter
is particularly severe, especially in the case of bursts originating
in dense stellar forming regions.
On the other hand photon--photon opacity may be an important
ingredient to shape the spectrum, and {\it the} reason why 
GRB spectra peak at around 300 keV.
As in the Brainerd model, the attractive feature is to link the energy
break to $m_ec^2$, while the difficulty is that all the primary
radiation emitted above $m_ec^2$ get absorbed.
Contrary to the Brainerd model, in this case the spectrum does not retain
its original slope above $\tau m_ec^2$, and implies that the GeV radiation
observed by EGRET for some GBRs is produced in the afterglow.
If the absorbed energy is not re--emitted, but remains in the form of lepton 
energy, this will significantly lower the efficiency of the burst to 
produce radiation.
On the other hand it is conceivable to expect that the created pairs
will radiate their energy in a short time, and that they can be even 
re--accelerated by the incoming fireball.
The net effect may be simply to increase the density of the
radiating particles, introducing a feedback process: an increased
density lowers the effective temperature $\to$ less energy is radiated
above $m_ec^2$ $\to$ the number of pairs produced via the ``mirror" 
process decreases $\to$ the new pair density decreases, and so on.
Another feedback is introduced by the fact that the photons
scattered back to the emitting shell will increase the number of
seed photons, softening the spectrum.
All these effects deserve a more detailed investigation, even if their 
time--dependence nature will made these studies quite complex.

\section{Pre--hypernova wind}

If the progenitors of GRBs are hypernovae (Paczy\'nski 1998), then we 
{\it must} consider the effects of the strong wind necessarily present
during the pre--hypernovae phase.
For illustration, assume  a mass loss 
$\dot m = 10^{-4}\dot m_{-4}\, \, M_\odot\,\, {\rm yr^{-1}}$ and a wind 
velocity $v=10^8 v_8$ cm s$^{-1}$.
The particle density $n_*$ near the surface of the pre--hypernova star 
of radius $R_*$ is 
\begin{equation}
n_* \, =\, {\dot m \over 4\pi R_*^2 v m_p} \, =\,
3.15 \times 10^{12} {\dot m_{-4} \over v_8 R_{*,12}^2}
\quad {\rm cm^{-3}}
\end{equation}
and scales as $(R/R_*)^{-2}$.
The mass contained in this wind decelerates the fireball at the 
deceleration radius $R_d$ where the wind mass equals the fireball mass
divided by $\Gamma$
\begin{equation}
{\dot m (R_d-R_*) \over v} = {E\over \Gamma^2 c^2} \, \to \,
R_d =  R_*+ {E v \over \Gamma^2\dot m c^2} = R_*+ 1.75\times 10^{13}
\, { v_8 E_{52} \over \Gamma_2^2 \dot m_{-4}} 
\,\, {\rm cm}
\end{equation}
As can be seen, the deceleration radius {\it is close to the transparency 
radius}, i.e. the distance at which the fireball becomes transparent. 
The first immediate consequence is that {\it internal shocks do not develop}.
The second immediate consequence is that the the optical depth of wind
material between $R_d$ and infinity is quite large
\begin{equation}
\tau_w(R_d-\infty)\, =\, \sigma_T\int_{R_d}^{\infty} n(R) dR\, \sim
\, 0.2 {\dot m_{-4} \over R_{d,13} v_8}
\end{equation}
Due to the $R^{-2}$ dependence of the density, most of the contribution
to this optical depth comes from material close to $R_d$.
Therefore all the effects discussed above (downscattering and pair 
reprocessing) would take place.

The conclusion is that the hypernova hypothesis implies a scenario
for the production of the burst and the afterglow quite different from
the internal/external shock scenario.
In this case in fact we have only external shocks between the fireball
and a very dense medium. 
The fireball {\it would decelerate} while producing the burst emission:
if a significant fraction of the bulk energy ends up into radiation,
this implies that the burst emission should contain more energy
than the time--integrated afterglow emission.
It may also imply a difference between the early and late burst emission,
due to the fact that the corresponding emitting zones may have
different $\Gamma$ factors.
Since this is not observed (see e.g. Fenimore 1999, this volume), 
this may be a problem for the hypernova idea, but since processes 
different from collisionless shocks may be operating (instabilities, 
turbulences and so on), this issue is worth investigating.

From the point of view of the radiation processes, the pre--hypernova
wind scenario offers an interesting possibility.
At early times, the heating--cooling balance (Eq. 1) gives sub or 
trans--relativistic lepton energies. 
As the energy density $U$ decreases, the cooling becomes less rapid and 
lepton energies increase, becoming relativistic.
Correspondingly, self--absorbed cyclo-synchrotron and multiple
Compton may originate the burst emission, while the afterglow
may correspond to synchrotron and self--Compton emission from 
relativistic electrons.
Transition from these two regimes may be smooth.
The self--absorbed cyclo--synchrotron radiation would increase its
relative importance as the Lorentz factor of the emitting leptons
increases, and develop a thin part when $\gamma$ becomes large enough
(e.g. $\gamma>100$), producing optical radiation.
This may be an alternative explanation for the prompt optical flash
of GRB990123.
At later times, the particle energy may reach its maximum possible
value (e.g. $\gamma=\Gamma m_p/m_e$), and then decrease following the 
usual prescriptions, with a power law time decay of the flux density.

\section{Conclusions}
If there is a balance between heating and cooling, the emitting leptons
reach typical energies which are mildly relativistic at most. 
The most efficient radiation process in this case is quasi--thermal
Comptonization of self--absorbed cyclo--synchrotron photons.
This process is characterized, in the quasi--saturated regime, by 
a spectrum which maintains its flat slope (in the power law part) even if
the emitting optical depth or the temperature change.
What changes is the relative importance of the Wien peak.
The emitting plasma may be dominated by the pairs produced through
photon--photon interactions in the high energy part of the spectrum,
and this may limit the effective temperature in a narrow range.
Most important, in this respect, is the exact shape of the high energy 
part of the particle distribution, which may differ from a pure Maxwellian.

The observed high energy cut--off of the burst emission is well defined,
and close to the rest--mass energy of the electron.
This fact is difficult to be explained both by ``standard" synchrotron
models and by Comptonization models.

It calls for a more robust interpretation, where the energy $m_ec^2$ enters
in a natural way.
We have argued that photon--photon absorption may play again a crucial
role if there is, in front of the fireball, some material scattering back
a fraction of the burst radiation.
This material may be the interstellar matter in a dense star forming region
or the matter blown out from a pre--hypernova star.
In the latter case the fireball is decelerated at typical distances
$R\sim 10^{13}$ cm, i.e. where it has become transparent.
There is no need to have internal shocks.
Other problems however arises, still to be investigated.

\end{document}